\documentclass[twocolumn,showpacs,preprintnumbers,amsmath,amssymb,floatfix]{revtex4}
\usepackage{epsf}
\usepackage{graphicx}
\usepackage{dcolumn}
\usepackage{bm}

\begin{document}

\title{Breathers and Thermal Relaxation in Fermi-Pasta-Ulam Arrays}

\author{Ramon Reigada}
\affiliation{Departament de Qu\'{\i}mica F\'{\i}sica, Universitat de
Barcelona, Avda.  Diagonal 647, 08028 Barcelona, Spain}

\author
{Antonio Sarmiento}
\affiliation{Instituto de Matem\'aticas, Universidad Nacional
Aut\'onoma de M\'exico, Ave.  Universidad s/n, 62200 Chamilpa,
Morelos, M\'exico} 

\author
{Katja Lindenberg}
\affiliation{Department of Chemistry and Biochemistry and Institute for
Nonlinear Science, University of California San Diego, 9500 Gilman Drive,
La Jolla, CA 92093-0340}
\date{\today}

\begin{abstract}
Breather stability and longevity in thermally relaxing nonlinear arrays
depend sensitively on their interactions with other excitations.  
We review the
relaxation of breathers in Fermi-Pasta-Ulam arrays, with a specific
focus on the different relaxation channels and their dependence on 
the interparticle interactions, dimensionality, initial condition, and
system parameters. 
\end{abstract}


\maketitle

{\bf Breathers are highly localized oscillatory excitations in
discrete nonlinear lattices that have been invoked as a possible way to
store and transport vibrational energy in a large variety of physical and
biophysical contexts.  A particular scenario where the robustness and
longevity of breathers has been a matter of considerable debate involves
nonlinear arrays subject to thermal relaxation via the connection of
surface sites to a cold environment.  The important questions are
these: can breathers (created spontaneously or by design) survive for
a long time in such a relaxing environment?   If they can survive, can they
move?  We detail answers to these questions, one of which is rather
unequivocal: breathers that move do not live very long. So is another:
breathers are quite robust when they do not move.  The more complicated
question then deals with the conditions that allow breathers
to remain stationary and undisturbed for a long time in a relaxing
environment.  We detail some conditions that lead to this outcome,
and others that definitely do not.}

\section{Introduction}

The localization of vibrational energy in discrete nonlinear arrays has 
attracted a huge amount of interest in the past several decades as a possible 
mechanism for the efficient storage and transport of energy (for recent
reviews see Refs.~\cite{FW,bri9} and references therein). 
More recently, the localization
and transport of 
vibrational energy has been invoked in a number of specific
physical settings including 
DNA~\cite{peyrard}, hydrocarbon structures~\cite{kopidakis}, the 
creation of vibrational intrinsic localized modes in anharmonic 
crystals~\cite{rossler}, photonic crystal waveguides~\cite{mingaleev}, and 
targeted energy transfer between donors and acceptors in 
biomolecules~\cite{directed}.  

Discrete nonlinear arrays in thermal equilibrium can support a variety
of stationary excitations; away from equilibrium stationarity may turn into
finite longevity,
and additional excitations may arise.  The possible excitations include
phonons associated with linear portions of the 
potential, solitons~\cite{kruskaletc,burlakovetc}
(long-wavelength excitations that persist from the continuum limit upon 
discretization), periodic 
breathers~\cite{FW,bri9,burlakovetc,kosevich,cretegny,bri2} 
(spatially localized time periodic excitations that persist from the 
anticontinuous limit upon coupling), and so-called chaotic 
breathers~\cite{cretegny} (localized excitations that evolve chaotically). 
Nonlinear excitations have been observed to arise
(spontaneously or by design) and survive for a long 
time in numerical experiments, and they clearly play an important role in 
determining the global macroscopic properties of nonlinear extended systems. 

Of particular interest to us is the dynamics of
breathers~\cite{we1d,we2d,weasymp}, a term that we
invoke rather loosely to denote an oscillatory excitation confined
to a very small number of adjacent lattice sites.  Since our interest 
lies in breathers as possible storers and carriers of
energy, we have concentrated on issues of longevity, and on lattices where
breathers can move most easily.  
Breathers are known to move more easily in nonlinear lattices with no
on-site interactions, and so we have focused on lattices with nonlinear
interactions.  Even more narrowly, herein we focus on the
nonequilibrium dynamics and relaxation of breathers in a typical
relaxation experiment where the surface of the system is 
connected to a cold (usually zero temperature) external thermal
reservoir.  We mostly (but not
exclusively) study one dimensional arrays, for which
the surface simply consists of the two end sites of a finite chain. 

We anticipate, and later detail, the following broad-brush description
of the relaxation of a breather whose energy is well above that of
phonon modes that may also be present in the nonlinear array.  When the
array boundaries are
connected to a zero-temperature heat bath, the breather will of course
eventually decay since the system must reach equilibrium at $T=0$. 
In other words, there is {\em necessarily}
leakage of energy out of the breather, although this process may 
in some cases be extremely slow.  A determinant limiter of breather
longevity is the
extreme sensitivity to collisions with long wavelength phonons and with
other localized excitations.  Such
collisions invariably contribute to the rapid degradation or breakup
of breathers into lower energy excitations.  Furthermore, collisions
with other excitations tend to set breathers in motion,
and motion in itself also contributes to energy leakage.
While breathers tend to decay rapidly
in the presence of long wavelength phonons and of other
nonlinear excitations, and are in this sense fragile,
{\em isolated}
breathers tend to remain stationary and to decay extremely slowly
and essentially exponentially over long time regimes, indicating a
single slow rate-limiting dominant contribution to the
intrinsic relaxation process. However, the particular values of
decay rates are strongly sensitive to particular conditions and
parameter values.  These statements will be made more quantitative below.

The organization of this paper is as follows. The model 
is presented in Sec.~\ref{model}, and a summary of the relaxation of
phonon modes in harmonic lattices~\cite{piazza}
is presented in Sec~\ref{linear}. In Sec.~\ref{nonlinear} we discuss the
relaxation behavior of a purely anharmonic lattice (no harmonic
interactions), that is, a relaxation scenario that involves {\em only}
nonlinear
excitations and no phonons.  Section~\ref{both} deals with breather
relaxation in arrays with both linear and nonlinear interactions, that is,
lattices that support phonons as well as nonlinear excitations.  
Finally, we present a summation of our findings in Sec.~\ref{concl}. 

\section{The Model}
\label{model}
Our model system in one dimension is described by the
Fermi-Pasta-Ulam (FPU) $\beta$-Hamiltonian
\begin{equation}
H = \sum_{i=1}^{N} \frac{\dot{x}_{i}^2}{2} + \sum_{i=1}^N
V(x_i-x_{i-1}),
\label{ham1}
\end{equation}
where $x_i$ is the displacement of particle $i$ from its equilibrium
position, $N$ is the number of sites, $V(z)$ is the FPU potential
\begin{equation}
V(z) = \frac{k}{2}z^2 +\frac{k'}{4}z^4,
\end{equation}
and $k$ and $k'$ are the harmonic
and anharmonic force constants, respectively.  The generalization to higher
dimensions is obvious. The relative values of the two constants can
be shifted by rescaling space and time.  In particular, by introducing new
variables $y_i=\alpha x_i$ and $\tau= t/\alpha$, where $\alpha$ is a
scaling constant, one finds that the scaled Hamiltonian
$\alpha^4 H$ in the new variables is again of
the form (\ref{ham1}) but with coupling constants $\alpha k$ and
$k'$.  The results are therefore related through appropriate
scaling for any choice of coupling constants {\em provided neither
is zero}. To cover all possible combinations of coupling
constants it is thus sufficient to consider only three distinct
cases:  $k'=0$ (harmonic), $k=0$ (purely anharmonic), and
$k=k'$ (mixed).  Throughout we assume free-end boundary conditions
($x_0=x_1$, $x_{N+1}=x_N$), and note that
although boundary
conditions do not strongly affect equilibrium properties, they {\em do}
affect relaxation dynamics. 

The equations of motion associated with the
Hamiltonian~(\ref{ham1}) are
\begin{equation}
\ddot{x_i} = -\frac{\partial}{\partial x_i} [V(x_i -x_{i-1}) +
V(x_{i+1}-x_i)].
\label{langzerot}
\end{equation}
In our subsequent discussion we consider a variety of initial conditions,
and observe the relaxation of the array to zero temperature when
the boundary sites are connected to a
zero-temperature environment by adding dissipation terms $-\gamma
\dot{x}_i$ to the equations of motion of these sites.  In one dimension the
boundary sites are $i=1$ and $i=N$.  The equations of motion
are integrated using a fourth order Runge-Kutta method with
time interval $\Delta t=5\times 10^{-4}$.  Further reduction leads to no
significant improvement.  Stability of the integration was checked for
isolated arrays: the energy remains constant to 10 or more significant
figures for all the cases and over all time ranges reported herein.

\section{Linear Modes}
\label{linear}
In the absence of anharmonic interactions the excitations of the system
are phonons whose behavior is well known.  It is useful to
briefly review this behavior because phonons may be present in the
nonlinear system, and their presence strongly affects the relaxation
behavior of nonlinear excitations.  

There are two informative measures to characterize the relaxation behavior
of an array initially thermalized at temperature $T$~\cite{thermalization}
and then
allowed to relax through the array boundaries into a zero temperature heat
bath~\cite{we1d,we2d}.  One is the total array energy as a
function of time, and the other
is the time dependent spectrum.  The total energy $\varepsilon(t)$ is
defined as the sum over symmetrized site energies, e.g. in one dimension 
\begin{eqnarray}
\varepsilon(t) &=& \sum_{i=1}^N E_i(t),\nonumber\\
E_i(t) &=& \frac{p_i^2}{2m} +
\frac{1}{2} V(x_{i+1}-x_i) + \frac{1}{2} V(x_i-x_{i-1}).
\end{eqnarray}
The time dependent spectrum is the Fourier transform of the
time dependent correlation function, 
\begin{equation}
S(\omega,t) = 2 \int_0^{\tau_{max}} C(\tau,t)\cos(\omega \tau) d\tau,
\end{equation}
where $\tau_{max} \equiv 2 \pi / \omega_{min}$ and $\omega_{min}$ is chosen
for a desired frequency resolution. The choice $\omega_{min} = 0.0982$
(corresponding to $\tau_{max}=64$), turns out to be numerically convenient.
In one dimension the time dependent correlation function is
\begin{eqnarray}
C(\tau,t)  &=& \frac{1}{(N-1)} \sum_{i=2}^{N}  \frac{1}{\Delta t}
\nonumber\\
&&\times \int_0^{\Delta t}  \left< \Delta_i(t - \tau ') \Delta_i(t - \tau'-
\tau) \right>  d\tau ',
\end{eqnarray}
where $\Delta_i=x_i - x_{i-1}$ is the relative displacement and
$\Delta t \equiv t_0 - \tau_{max}$.
The correlation function is thus
an average over the interval $[t-t_0,t]$, and $t_0$ is a time interval
chosen to be
short enough for the correlation function not to change appreciably but
long enough for statistical purposes.  In our work we take $t_0 = 100$. 
The generalization of these definitions to higher dimensions is
straightforward.

\begin{figure}[htb]
\begin{center}
\leavevmode
\epsfxsize = 3.2in
\epsffile{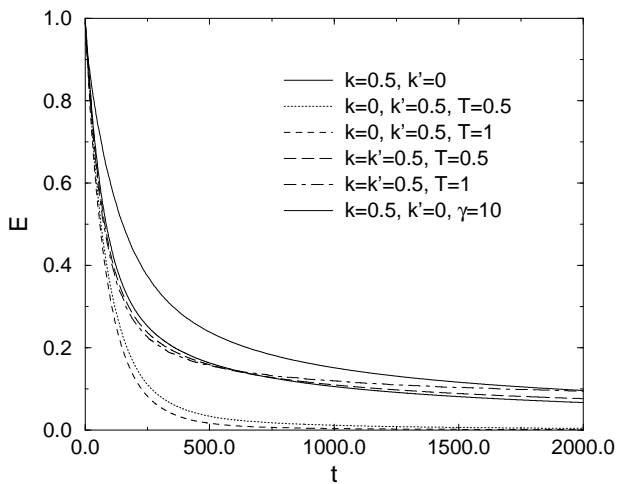}
\end{center}
\caption{Normalized energy vs. time for various one dimensional relaxing
arrays with $N=50$. Initially each array is in thermal equilibrium at the
temperature indicated in the figure. The normalized energy of the
harmonic array is independent of temperature.
In all cases except for the thin solid line.
$\gamma=0.1$.  The thin solid curve is for the harmonic chain
with $\gamma=10$.}
\label{figure1}
\end{figure}

If the chain is sufficiently long and the damping constant $\gamma$ at the
ends sufficiently small, then the phonon dynamics in the array are not
greatly disturbed by the damping, and the relaxation process is 
perturbative.  The two principal characteristics of
phonon relaxation are
then that (1) phonons of each frequency relax independently of other
phonons, and (2) the relaxation times are wavevector dependent.  The
relaxation time for phonons of wavevector $q$ in the small damping limit
has been calculated
by Piazza et al.~\cite{piazza} for different boundary conditions.  For
free-end boundaries they obtain the decay times in one dimension
$\tau(q)=\tau_0/\cos^2(q/2)$, where $q=n\pi/N$, $n=0,1,\cdots,N-1$ are the
allowed wavevectors, and $\tau_0=N/2\gamma$.  Long wavelength
phonons thus  decay more rapidly [$\tau(q)\sim{\mathcal O}(N/\gamma)$]
than do band edge phonons [$\tau(q)\sim {\mathcal O}(N^3/\gamma)$].  
The associated normalized chain energy $E(t)$ as a function of time can be
evaluated exactly in this limit:  
\begin{eqnarray}
E(t)\equiv\frac{\varepsilon(t)}{\varepsilon(0)}&=&\frac{1}{\pi}\int_0^\pi
e^{-2t/\tau(q)} dq = e^{-t/{\tau_0}} I_0(t/\tau_0) \nonumber \\
&=&e^{-t/\tau_0} \qquad \quad \hbox{for $t \ll \tau_0$},
\nonumber\\
&=&\left(\frac{2 \pi t}{\tau_0} \right)^{-1/2} 
\hbox{for $t \gg \tau_0$}.
\label{energy1}
\end{eqnarray}
Here $I_0$ is the modified zero-order Bessel function.  The short time
exponential behavior reflects the earliest decay of the long wavelength
phonons.  The power law behavior at long times arises from the cascade of
relaxation times that contribute to the process.  In a finite
chain at very long times the decay will revert to exponential when only
the shortest wavelength phonons remain, with a characteristic decay time
of ${\mathcal O}(N^3/\gamma)$.  Similar arguments are immediately
applicable in higher dimensions. 
Note that the energy decay in the harmonic arrays is {\em independent of
the initial temperature}.
The decay curve for a one dimensional harmonic chain with $\gamma=0.1$ is
shown in Fig.~\ref{figure1}.  A typical time-dependent
landscape in which energy magnitudes are represented by varying gray
scales has been presented in our earlier work~\cite{we1d,we2d}.

\begin{figure}[htb]
\begin{center}
\leavevmode
\epsfxsize = 3.2in
\epsffile{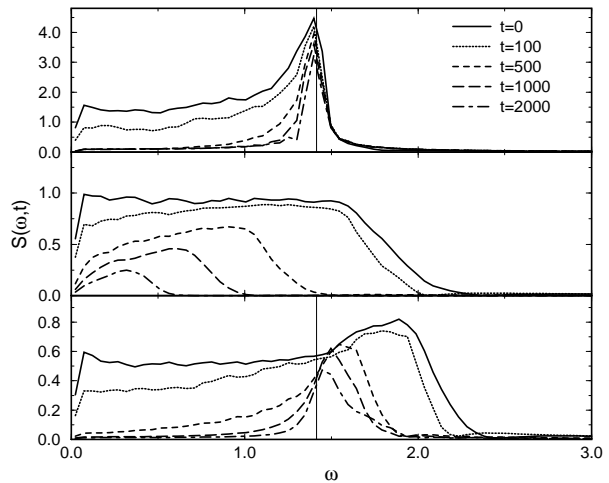}
\end{center}
\caption{
Time evolution of spectra for various relaxing 1d arrays of $50$ sites
relaxing from a thermalized initial condition. In all cases the
initial temperature is $T=0.5$, and $\gamma=0.1$. 
First panel: harmonic interactions ($k = 0.5$);
second panel: purely anharmonic interactions ($k' = 0.5$); third panel: mixed
interactions ($k = k' = 0.5$).  The thin vertical lines
indicate the harmonic frequency $\omega = \sqrt{4k} = \sqrt{2}$.}
\label{figure2}
\end{figure}

It should be recognized that the results reported above are restricted to
{\em weak} damping.  If the damping coefficient becomes large
($\gamma \gg \sqrt{4k}=\sqrt{2}$), the theory of Piazza et al.
no longer applies.  The
boundaries act more like hard walls and the phonon decay slows down with
increasing $\gamma$. As an example we have included the 1d curve for
$\gamma=10$ in Fig.~\ref{figure1}, which is clearly qualitatively similar
to  the $\gamma=0.1$ curve. 
Although not shown, we note that the
decay curves of the normalized energies for the damping parameters
$\gamma=0.01$ and $\gamma=100$ are also quite similar to one another.
Note that the relaxation {\em slows down} with increasing damping.
The concept of ``optimal damping" that this behavior implies is intriguing. 

\begin{figure}[htb]
\begin{center}
\leavevmode
\epsfxsize = 3.2in
\epsffile{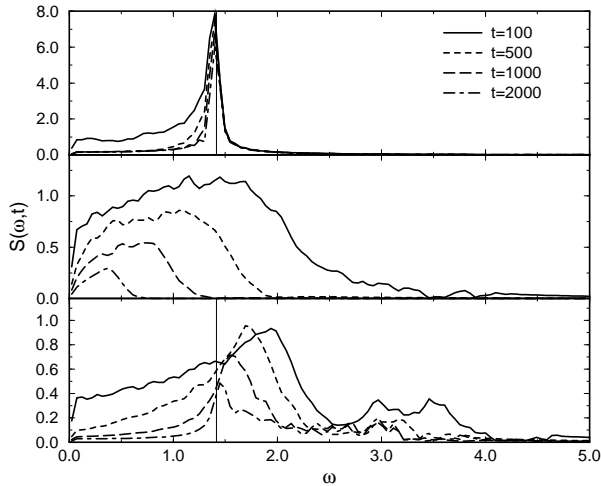}
\end{center}
\caption{
Time evolution of spectra for various relaxing 1d arrays initially at
$T=0.1$.  At $t=0$ a high amplitude localized excitation is injected near
the center of the chain.  In all cases $\gamma=0.1$. 
First panel: harmonic interactions ($k = 0.5$);
second panel: purely anharmonic interactions ($k' = 0.5$); third panel: mixed
interactions ($k = k' = 0.5$).  The thin vertical again lines
indicate the harmonic frequency $\omega = \sqrt{4k} = \sqrt{2}$.}
\label{figure3}
\end{figure}

The spectral progression of the phonon-by-phonon decay in the weak damping
case is interesting because it provides a forum to differentiate linear
from nonlinear systems in this limit.
The first panel of Fig.~\ref{figure2} shows the time progression of the
spectrum of an initially thermalized harmonic chain as it relaxes.
The $t=0$ curve is the
equilibrium spectrum of the harmonic chain at temperature $T$ and
can be calculated analytically~\cite{ourappendix}.  We note that the only
temperature dependence of this spectrum is an overall coefficient $T$.
The progressive relaxation starting from the lower part of the
spectrum and moving upward is clearly evident; by
time $t=2000$ only the highest frequency phonons survive.  The progression
in 2d is very similar.

Although the above analysis started from thermalized arrays, 
similar conclusions apply no matter the initial
condition (including highly localized excitations), and no matter if
and when additional excitations are injected in the array~\cite{we1d,we2d}. 
Any distribution of energy consists of a
superposition of phonons,
and each phonon relaxes independently 
with its own characteristic decay rate.  This is consistent with the
spectral progression shown in the first panel of
Fig.~\ref{figure3}.  Here an odd parity localized excitation of amplitudes 
$-A/2$, $A$, $-A/2$ on three successive sites 
has been injected near the center of the thermalized chain at $t=0$. 
As before, the 
phonons decay progressively starting from the lower portion of the spectrum.
The detailed $E(t)$ curve would of course be modified because the
distribution of energy among the phonon modes is now different.  

\section{Nonlinear Modes}
\label{nonlinear}
As a second ``extreme" case we consider a purely anharmonic array ($k=0$).
In the absence of harmonic forces the system supports {\em no phonons},
a condition
that has been referred to as a {\em sonic vacuum}~\cite{nesterenko}.
This absence leads to a
dynamic and relaxational behavior quite distinct from that observed
in a mixed array (next section), where phonons do constitute part of
the spectrum.  The precise nature of the full spectrum of excitations
in this array is not known, but it certainly includes highly
localized excitations.  Indeed,
breather solutions are exact in such chains when the power of the anharmonic 
potential (which here is $4$) as well as $N$ go to
infinity~\cite{breatherF}.

Consider first the relaxation of a purely anharmonic 1d array initially
thermalized at temperature $T$~\cite{we1d}.  Although we do not study it
here, we note that the temperature
dependence of the relaxation process is more complex than in the
harmonic array because now the frequencies of excitations depend very
markedly on their energy.  
A characteristic of the purely anharmonic 1d array at any temperature
is the essentially strictly exponential tail of the normalized energy
$E(t)$. For example, the exponential decay of the purely anharmonic 
curves shown in Fig.~\ref{figure1} has been ascertained in detail in
Ref.~\cite{we1d}.  This behavior
implies that the purely anharmonic array approaches its new equilibrium much
more rapidly than a harmonic system, and is indicative of a single
predominant rate-limiting decay channel.   Furthermore, the
short time relaxation is more rapid with increasing temperature. 

The spectral progression of the relaxation
process is illustrated in the second panel of Fig.~\ref{figure2}.  
Unlike the harmonic case, the equilibrium
spectrum of the anharmonic chain broadens with increasing temperature
because higher energy excitations involve higher frequencies.
The higher frequency portions of the spectrum are
observed to decay first, exactly opposite to the harmonic chain. 
We have found that the relaxation pathway is for
the high frequency portions of the spectrum to degrade rapidly into lower
frequency excitations~\cite{we1d}; such a degradation pathway
is possible here
since individual frequencies are not associated with normal modes in the
anharmonic system. In turn, these lower frequency excitations decay into the
reservoir through the ends of the chain.
The high frequency components of the spectrum are mainly
associated with
mobile localized modes that degrade into lower energy excitations (often
also localized and mobile) as they
move and collide with one another, and this degradation occurs 
relatively quickly.
It is well known that higher frequency and/or higher amplitude
localized modes can move at higher
velocities~\cite{kosevich,cretegny,bourbonnais}. It is also known that
while in motion such modes lose energy through collisions with other
excitations.  This picture is thus consistent with the more rapid decay of
$E(t)$ with increasing temperature.
The lower frequency excitations are in turn absorbed into
the cold reservoir, but continue to be replenished through the degradation
process.
The decay of low frequency excitations into the reservoir dominate
the exponential
tail in the decay curves such as those
seen in Fig.~\ref{figure1}.  We stress that
this discussion only covers the predominant relaxation mechanisms.
A concurrent direct but slow relaxation of high
frequency excitations through the boundary sites may also take place.  For
example, when a highly mobile localized excitation reaches a boundary,
it typically remains at the boundary for about one period of
oscillation (which is short for a highly energetic excitation), during
which it loses a small portion of its energy to the reservoir. The
remaining excitation is reflected back into the
chain, where it will continue to lose energy through other collisions
and/or re-arrival at the boundaries. 

The relaxation dynamics of the purely anharmonic array in 2d differs from
the one-dimensional case in a number of ways~\cite{we2d}.
Mainly, the energy decay is
considerably slower because localized excitations in 2d are not nearly as mobile
as in 1d; consequently, the energy loss caused by motion and by
collisions is slowed down.   Furthermore, the decay is no longer
exponential.  Indeed, whereas in one dimension the degradation process
of higher frequency excitations into lower frequency ones
is faster than the decay of low frequency excitations into the reservoir,
in two dimensions this is no longer the case.   This leads
to spectral bottlenecks and competing time scales. 
We also find that increasing the array size leads to slower
degradation of the high-frequency components and to more pronounced
spectral bottlenecks in the mid-frequency range.  Still, with 
increasing initial temperature the total system energy decays more
rapidly, which is
consistent with our assertion that mobility (low as it may be) in the
purely hard arrays increases with energy and hence with increasing 
initial temperature. 
We thus conclude that although breathers may be among the thermal
excitations in purely nonlinear thermalized arrays, their rapid
degradation makes it difficult to identify their precise dynamics.  

In order to specifically focus on breather dynamics and the effects of
other excitations on breathers, we 
inject a high amplitude localized excitation (one whose energy is
much greater than
those of the thermal excitations) at time
$t=0$ and observe its evolution. Explicitly, we create an ``odd parity"
excitation with amplitudes $-A/2$, $A$, $-A/2$ on three successive
sites away from the chain boundaries.  Note that this is not an exact
breather for this array, so some amplitude re-accommodation
accompanied by some energy shedding necessarily takes place.
The ``bottom line" of this experiment is that the resulting breather
is extremely robust {\em in isolation}, but extremely fragile
when disturbed in any way.  To see this, consider the case of such an
excitation injected into a
purely nonlinear array that has first been thermalized to temperature $T$.
Although the
details vary somewhat from one realization to
another, it invariably happens very quickly that the other excitations in
the medium set the breather in motion, and it loses energy
mainly through its collisions with other excitations.  A detailed
observation of trajectories shows that after a short time the
injected excitation begins to move in one direction or the other with equal
probability, and continues
moving for a period of random duration, during
which it loses energy. The excitation stops moving for a random period of
time, until it is again set in motion in either direction for
another random period of time. While stationary, the excitation has even
parity, but when it moves it alternates between even and odd parity.
The collision partners in this sequence of events are principally the
persistent low-frequency excitations described earlier in the thermal
relaxation process~\cite{we1d}. Even while its amplitude and consequently
its characteristic frequency are decreasing, the breather
remains highly localized (over essentially three or four sites) throughout
its lifetime.  The spectral progression of the relaxation
process is illustrated in the second panel of Fig.~\ref{figure3}.  
In accordance with our description, it does not differ significantly
from that of Fig.~\ref{figure2}.

We have not measured the lifetime of the breather in this scenario, but 
note that it is much shorter than lifetimes to be reported later in
mixed arrays, and
varies quite a bit from one realization to another.  It is, however,
possible to follow the breather until it disappears, and to
characterize its motion in terms of the mean square displacement
$\left< x^2(t)\right>$ of its
center from its initial location.  In 1d we find superdiffusive
motion, $\left< x^2(t)\right> \sim
t^{1.5}$, over the entire lifetime of the excitation and for a variety of
values of the coupling constant $k'$, excitation amplitude, and
temperature~\cite{we1d,we2d}.  Parameter
variations seem only to affect the prefactor in this relation.  It does
not even matter {\em when} in the course of
the relaxation process the localized excitation is introduced: its mean
squared displacement grows with the same exponent, $1.5$, until it
is extinguished into the background. This corroborates the role of
the persistent low-frequency excitations.

The situation is somewhat different in a purely anharmonic 2d array.
Here we inject an excitation of amplitude $A$ at a central site and $-A/4$
at each of the four nearest neighbor sites.  It is
more difficult in 2d to set a breather in motion since it is more
difficult to bring about the symmetry breaking behavior that favors such
motion (even parity breathers do not exist in 2d). This observation is
consistent with our earlier comment concerning lower mobilities of breathers
that may arise
spontaneously in thermalized 2d arrays.  Lower breather mobility
of course does not
preclude collisions with other mobile excitations that lead to energy loss.
In any case, the breather survives
for a longer time than it does in 1d, and its mean square displacement is
subdiffusive,  $\left< r^2(t)\right> \sim t^{0.89}$.  

\begin{figure}[htb]
\begin{center}
\leavevmode
\epsfxsize = 3.2in
\epsffile{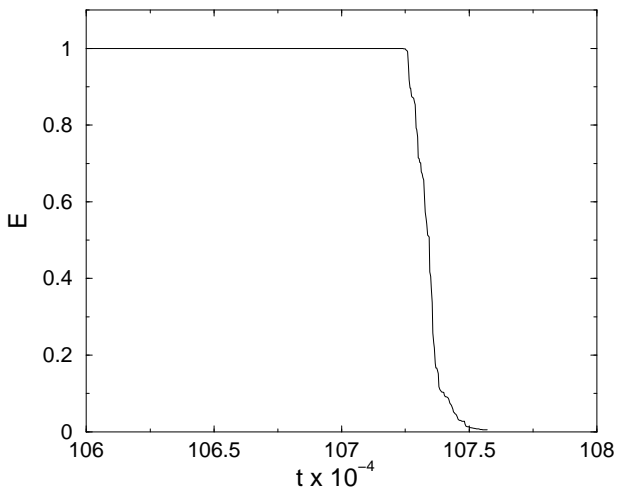}
\end{center}
\caption{Normalized energy for a purely anharmonic array ($k'=0.5$)
of 15 sites in which a breather of amplitude $A=0.5$ has been injected
into a zero temperature chain at $t=0$.  The damping parameter
$\gamma=1$.}
\label{figure4}
\end{figure}

The scenario changes dramatically when the localized excitation is
injected into a
purely nonlinear chain {\em that is initially at zero temperature}.  
Breathers are almost exact stationary solutions
in very long purely anharmonic chains regardless of the
value of the damping parameter at
the ends.  In our simulations we inject
an excitation of amplitude $A=0.5$ near the chain center at time $t=0$ in a
chain of $N=31$ sites.
For damping parameters $\gamma=0.01, 1$, and $100$ the energy remains
essentially unchanged to all the controlled significant figures ($12$) 
for the duration of our simulation, $t=3\times
10^6$. We can also follow the motion of the sites surrounding the
breather, and find that sites other than the three involved in the
breather motion are almost stationary.  Of course there {\em must}
be an energy
leakage out of the chain since equilibrium must eventually be reached, but 
for the chain length and simulation
times in this particular experiment it was not
discernible, so the decay is extremely slow.
However, this longevity is, as before, fragile in that it is
quickly destroyed by practically any perturbation.
The source and nature
of the perturbation does not much matter.  A perturbation might
arise, for example, when the slowly leaking energy reaches the chain
boundaries and is partially reflected back toward the breather.  This
finite size effect should be observable if we shorten the chain
sufficiently.  For a chain of $N=17$ sites we still see no decay, but
when we further shorten the chain to $N=15$ sites the breather
survives almost undisturbed until a
time of ${\mathcal O}(10^6)$ and then disintegrates rapidly, over a time
scale of a few thousand time units.  This is illustrated for one
realization in Fig.~\ref{figure4}.  When $N=13$ the breather survives
undisturbed for only about $7\times 10^3$ time units and the
decay occurs over a time period of about $10^3$.  

Our conclusion is thus that breathers are exceptionally stable in purely
anharmonic arrays relaxing into a zero temperature heat bath through the
array boundaries {\em as long as there is no perturbation of any kind
near the breather}.  Any perturbation, including even the smallest thermal
perturbations or perturbations that reach the breather from the system
boundaries, causes a rapid degradation of the breather.

\section{Mixed Arrays}
\label{both}
Mixed arrays, that is, arrays with both harmonic and anharmonic
interactions, are the most versatile because they support both linear
and nonlinear modes. The interplay of the two introduces new effects
in the breather relaxation problem and a greater variability in breather
dynamics.  Also, the mixed array is representative of a larger variety of
physical situations than the purely anharmonic.  
Furthermore, we will see that breathers
may be exceptionally stable in relaxing mixed arrays.  Therefore, a study
of the full time evolution of the system during the relaxation process
requires longer time histories than in the previous cases.  Our
discussion in this section deals only
with 1d arrays because the short time behavior in 2d is very similar to
that of the 1d systems, and because we have not carried out a long time
study for the 2d system.

Again we begin with the relaxation of an array initially
thermalized at temperature $T$~\cite{we1d}.  At low temperatures ($T$
smaller than the phonon bandwidth associated with the linear portion of
the potential) the relaxation is essentially identical to that of a
harmonic chain.  Interesting behavior clearly requires higher temperatures
so that nonlinear excitations well above the phonon band can be part of
the thermal mix.  In Figs.~\ref{figure1} and \ref{figure2} we
observe the early time behavior of the energy relaxation and the spectral
progression of the relaxation.  The energy is at first seen to decay
more rapidly than in the harmonic array.  This is a consequence of the
presence of both low-frequency phonons
{\em and} high-frequency excitations in the system. Energy relaxation
and decay thus involves {\em both} of the mechanisms discussed earlier,
namely, that which characterizes the relaxation of the harmonic chain
and, concurrently, that which characterizes the purely anharmonic chain.
Again, because initially the high-frequency modes move more rapidly
at higher temperatures, the early time decay is faster with increasing 
initial temperatures.  That both low and high frequency modes relax rapidly
is clearly seen in the third panel of
Fig.~\ref{figure2}, which quickly loses both low (as in the first panel)
and high (as in the second panel) frequency
portions of the spectrum.  In the energy decay curve there is then a
crossing after which the mixed chain relaxes much more {\em slowly} than
the harmonic and the purely anharmonic.  This occurs when the low
frequency modes (phonons) have essentially all decayed, and only certain
high-frequency spectral components remain, as clearly seen in the
spectrum.   

\begin{figure}[htb]
\begin{center}
\leavevmode
\epsfxsize = 2.5in
\epsffile{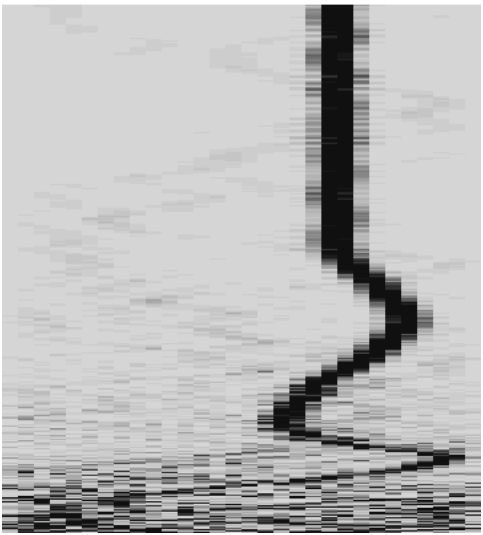}
\caption{
Energy landscape of 30-site mixed array initially thermalized at $T=0.5$.
Other parameters: $k=k'=0.5$, $\gamma=0.1$.
Time advances along the $y$-axis until $t=1000$.
A gray scale is used to represent the local energy,
with darker shading corresponding to more energetic regions. 
}
\label{figure5}
\end{center}
\end{figure}

The excitations that remain after this initial shedding include, with some
probability that depends on temperature (but not with certainty),
spontaneously created
quasi-stationary breathers that decay {\em extremely}
slowly.  It is a noteworthy reminder that short wavelength zone-boundary
phonons that relax much more slowly than long wavelength phonons actually
contribute to the spontaneous formation of breathers (``modulational
instability")~\cite{daumont}.  The decay of the breather
is slower with increasing temperature because the spontaneously
created breathers are more energetic.  A typical realization in which a
breather appears spontaneously is shown in
Fig.~\ref{figure5}.
In stark contrast with the purely anharmonic array, there is no sonic
vacuum in the mixed array.  The harmonic contribution to the
interaction, which provides the phonon excitations, allows the relaxation
process to ``sweep" the system clean of the excitations that most readily
perturb the breather, and thus makes it possible for the breather to persist
in the relaxing environment. 

\begin{figure}[htb]
\begin{center}
\epsfxsize = 3.2in
\epsffile{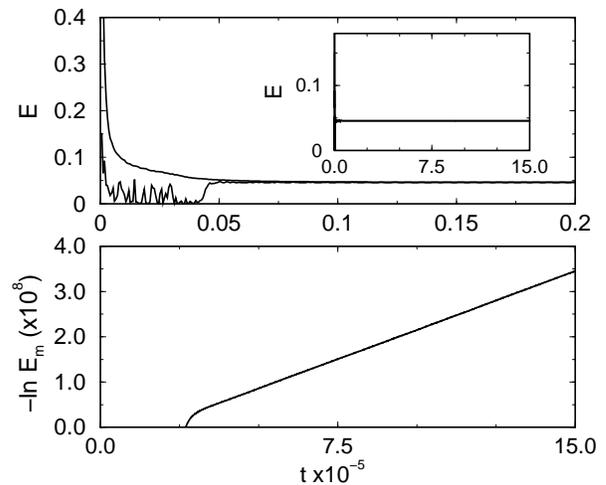}
\end{center}
\caption{
Upper panel: the smooth curve is the normalized energy as a
function of time for a chain of $30$ sites initially thermalized at $T=0.5$
and connected through its ends to a zero-temperature heat bath
($\gamma=0.1$).  The initially jagged curve is the
normalized energy on sites $13$, $14$, $15$, and $16$. The inset shows the
temporal evolution of the energy on these four sites
over a longer time scale. Lower panel:
$\left[-\ln E_m(t)\right]$ vs $t$ for the same chain.
}
\label{figure6}
\end{figure}

Next we examine in detail the continued evolution of a spontaneously
created breather in our relaxing environment~\cite{weasymp}, going to
much longer times in our simulations than those reported so far in order
to ascertain the relaxation behavior in more detail.  We
distinguish the
normalized energy $E(t)$ [cf. Eq.~(\ref{energy1})] from the
modified normalized energy $E_m(t)$,
\begin{equation}
E(t) = \frac{\varepsilon(t)}{\varepsilon(0)}, \qquad
E_m(t)=\frac{\varepsilon(t)}{\varepsilon(\tau_m)}.
\label{normalizede}
\end{equation}
The denominator in the first contains the initial energies,
and in the second the energies after the
discarded low energy excitations (including all phonons) have dissipated,
but before the remaining breather has decayed appreciably. In our
simulations we take $\tau_m=40,000$.
Figure~\ref{figure6} shows the evolution of the normalized energy in a
chain of $30$ sites initially thermalized at temperature $T=0.5$,
as well as the energy
in only the four sites $i=13, 14, 15, 16$.  After a relatively
short time ($5000$ time units in this particular realization) almost all
of the energy settles in these sites and remains there.
The excitation around the four sites turns out
to be an ``even parity" breather, with maximum displacements $A$ and $-A$
alternating on sites $14$ and $15$, smaller but not negligible amplitudes
at sites $13$ and $16$, and essentially no motion at the other sites.
The breather is coincidentally at the center of the chain, but may appear
anywhere in different realizations (cf. Fig.~\ref{figure5}),
particularly in longer chains.
The frequency of the breather, initially $\omega=1.633$,
decreases very little for the duration of the simulation.
We also present the modified normalized energy,
whose decay is clearly exponential over the times shown, with
an enormously long time constant, $\tau=3.5\times 10^{13}$.
Thus this breather, even in our relatively short chain, is
essentially stationary.  

In order to ascertain our breather stability scenario and the role of 
phonons in the stability, we
explicitly inject a breather into a chain 
that is in thermal equilibrium at a very
low temperature (low in the sense that the
spontaneous formation of breathers is highly unlikely).
The chain is then allowed to relax into a zero
temperature heat bath (cf. third panel in Fig.~\ref{figure3}). 
As expected,
we find that the thermal background invariably
sets the breather in motion, and causes the breather to collide with other
excitations and with the chain boundaries.  The resulting
decay of the breather is then in general much faster than
in the scenario where the breather is created spontaneously during the
relaxation process (and certainly much faster 
than that of a breather of the same amplitude injected into
a zero temperature chain, cf. below).  We find this behavior
even when the temperature is extremely low.  For example, 
a breather of initial amplitude $A=0.5$ injected into a
chain thermalized at $T=10^{-6}$ has a decay time of
$\tau=1.3\times 10^6$ (the value varies from one realization
to another, but not by much).  With an initial amplitude of
$A=0.55$ we find $\tau=2.1\times 10^6$. 
Note that it does not much
matter whether the injected breather is of even or odd parity (here we
have injected an odd one)~\cite{page}. 
As we will see below, although the initial
temperature is extremely low, these decay times are orders
of magnitude shorter than those of a breather injected in a zero
temperature chain, the crucial difference being the presence of low
wavelength phonons in the former but not in the latter.
To support this description we have also observed a breather in a
chain connected to a heat bath that is maintained at an
extremely low but nonzero temperature.  The breather in this case is always
fragile, continuing to move and lose energy rapidly until it
degrades completely.

We now move on to our second scenario, namely, we inject the
excitation into a chain at {\em zero} temperature~\cite{weasymp}.
Although we have not done so in detail in the cases presented so far, here
we have carried out a more detailed analysis of dependences of the
breather dynamics and relaxation on a number of parameters.
The results are revealing and, in some cases,
somewhat unexpected.   In our first ``experiment" 
we create an odd-parity excitation of amplitude $A$ exactly in
the middle of a chain of $N=31$ sites, and set the end-site dissipation
parameter $\gamma=1$.  The breather discards some energy that travels
toward the chain ends and dissipates quickly, in the time $\tau_m$ [cf.
Eq.~(\ref{normalizede})], across
the ends of the chain.  The remaining energy stays localized in the middle
of the chain, most of it ($98\%$ for $A=0.5$) on the three initially
excited sites, and decays exponentially
with an extremely long decay time $\tau$.

\begin{figure}[htb]
\begin{center}
\epsfxsize = 3.2in
\epsffile{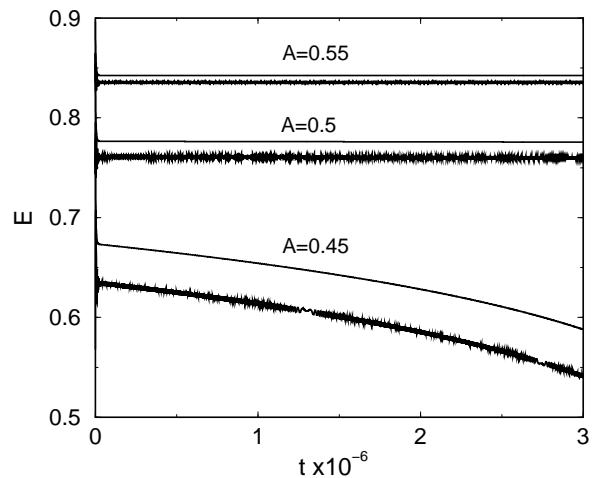}
\end{center}
\caption{Decay of the
normalized energy $E(t)$ for three different values
of the initial amplitude $A$ for chains of $N=31$ sites connected at the
ends to a zero-temperature bath.  The odd-parity
breather is injected at the center of
the chain.  The dissipation parameter $\gamma=1$. The
thin lines represent the total energy remaining in the chain, and the bold
lines the portion of the remaining energy that is localized on the three
initially excited sites.
}
\label{figure7}
\end{figure}

\begin{figure}[htb]
\begin{center}
\epsfxsize = 3.2in
\epsffile{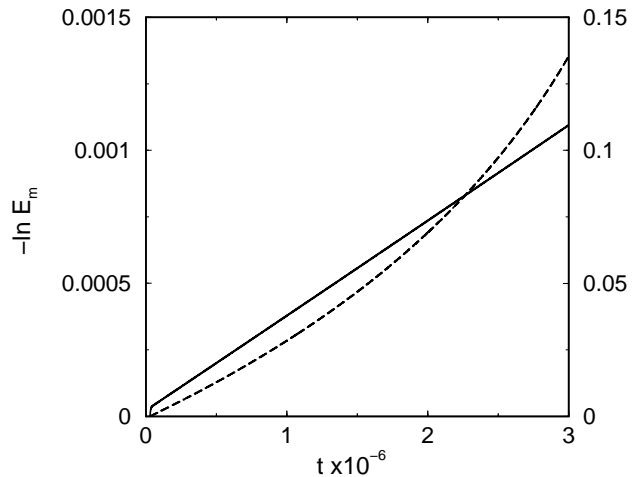}
\end{center}
\caption{
$\left[-\ln E_m(t)\right]$ vs $t$ for two initial
amplitudes, $A=0.5$ (solid curve, left scale)
and $A=0.45$ (dashed curve, right scale), for a chain of $31$ sites
with an odd-parity breather at the center and $\gamma=1$.
}
\label{figure8}
\end{figure}

In Fig.~\ref{figure7} we show typical results for $E(t)$
for three excitation amplitudes over more than six decades of time.
The modified normalized energy follows essentially the same behavior.
If the decay of the long-lived excitation is exponential, we expect
$\left[-\ln E(t)\right]$ and $\left[-\ln E_m(t)\right]$
vs $t$ to be straight
lines over appropriately long time intervals. In Fig.~\ref{figure8} we
clearly see this behavior, which extends over the entire
simulation time interval for
the higher amplitude excitation.
The slope for the $A=0.5$ curve
leads to a decay time of $\tau=2.8\times 10^{9}$, a specific number
reported here principally to stress its enormous magnitude compared to
phonon relaxation times. 
The change in slope of the curve associated with
the lower amplitude breather
captures the slow change in the decay rate as the breather
frequency edges toward the phonon band.  Here we also see clearly that
the more energetic breather relaxes more slowly.

\begin{figure}[htb]
\begin{center}
\epsfxsize = 3.2in
\epsffile{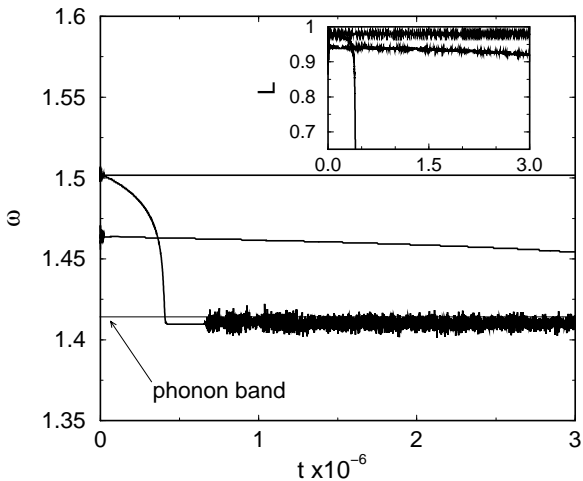}
\end{center}
\caption{Breather frequency as a function of time. 
Curve that persists at
the highest frequency ($\omega\sim 1.5$):
$A=0.5$, $N=31$. Curve that begins at $\omega\sim 1.465$ and decreases
gently: $A=0.45$, $N=31$. Curve that turns sharply downward: $A=0.5$,
$N=21$.  Inset: associated localization parameters in the same order.
}
\label{figure9}
\end{figure}

A breather of a given amplitude has a characteristic predominant
frequency.  In Fig.~\ref{figure9} we show this frequency in relation to the
phonon band edge as a function of time for various cases.
For 31-site chains, the 
frequency of the breather of initial amplitude $A=0.5$ decreases very
little
over the entire simulation, while that of initial amplitude
$A=0.45$ decreases more markedly.  Consistent with the fact that the
breather does not disappear entirely in the time range shown,
its frequency never reaches
the phonon band edge.  If the initial amplitude of the excitation is
sufficiently low,
or the simulation time sufficiently long, or the chain sufficiently
short, the
breather is seen to disappear. This last case is illustrated in
the figure for a breather of initial amplitude $A=0.5$ in a 21-site chain.
The breather disintegrates entirely when
its frequency reaches the phonon band edge.  The inset
shows $L$, the ratio of the energy of the five sites centered on the
breather to the total energy.  $L$ is of order unity when
most of the energy is localized on a small number of sites.
Note that the lifetime of this breather, which
is of $O(5\times 10^5)$, is still
much longer than the longest phonon lifetime, which is of $O(10^4)$.

The above results are typical of one particular set of conditions: a
breather created exactly in the middle of a chain of $N$ sites whose
ends are connected to a zero-temperature bath with dissipation parameter
$\gamma=1$.  It is interesting to explore the consequences of changing
some of these conditions.  We find that
the dependence of the chain energy relaxation times on the initial
amplitude of the breather is monotonic and decreases sharply with
decreasing breather amplitude.  Over a simulation time of $3\times 10^6$
we find that a breather of initial amplitude $A=0.6$ decays exponentially
with a time constant $\tau=2.8\times 10^{14}$.  For amplitude $A=0.5$ 
we find $\tau=2.8\times 10^9$, and for $A=0.45$ the decay is no longer
{\em strictly} exponential, decreasing slightly from $3.0\times 10^7$ to
$1.5\times 10^7$ over the course of the simulation. 

\begin{figure}[htb]
\begin{center}
\epsfxsize = 2.in
\epsffile{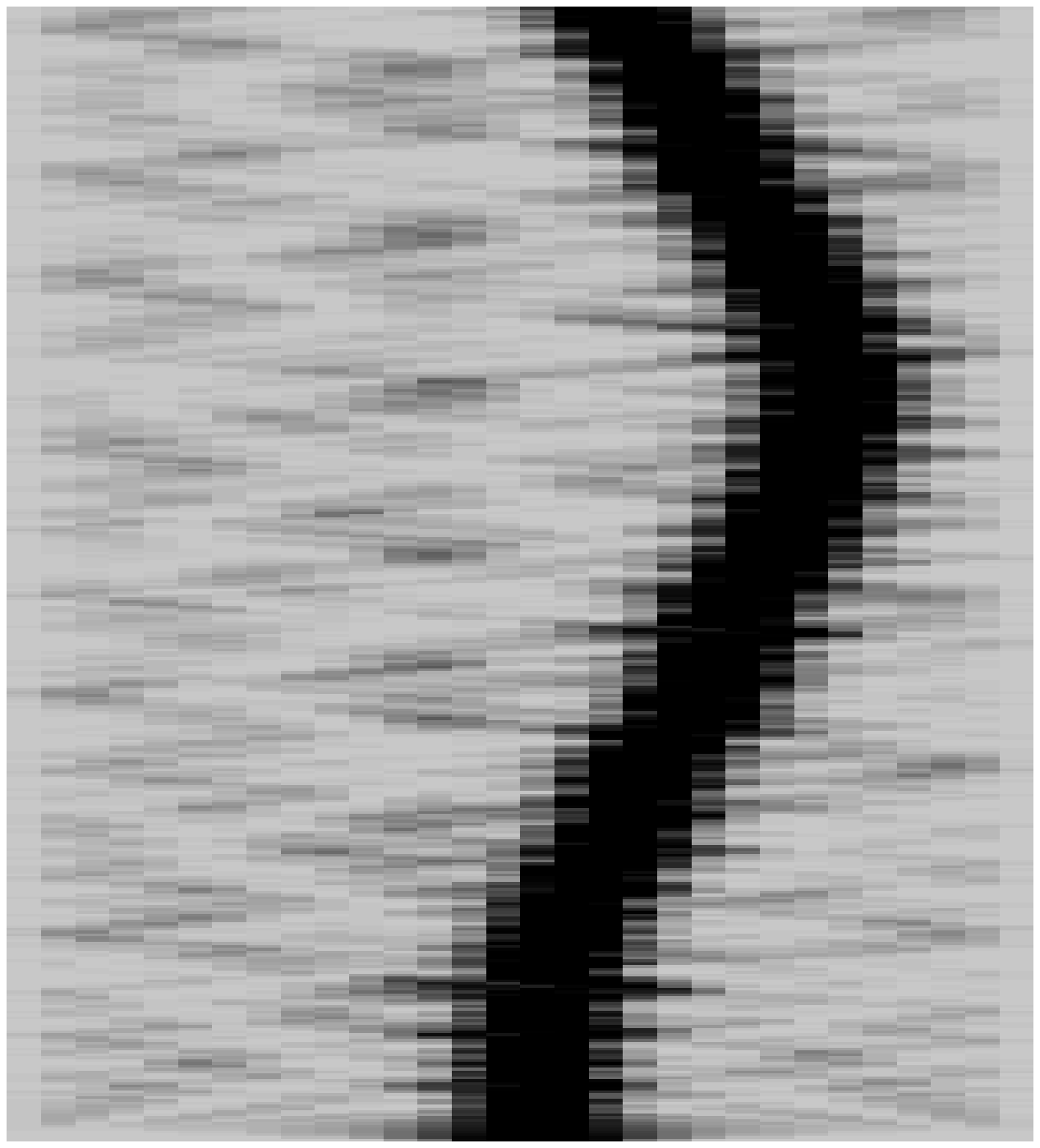}
\epsfxsize = 2.in
\epsffile{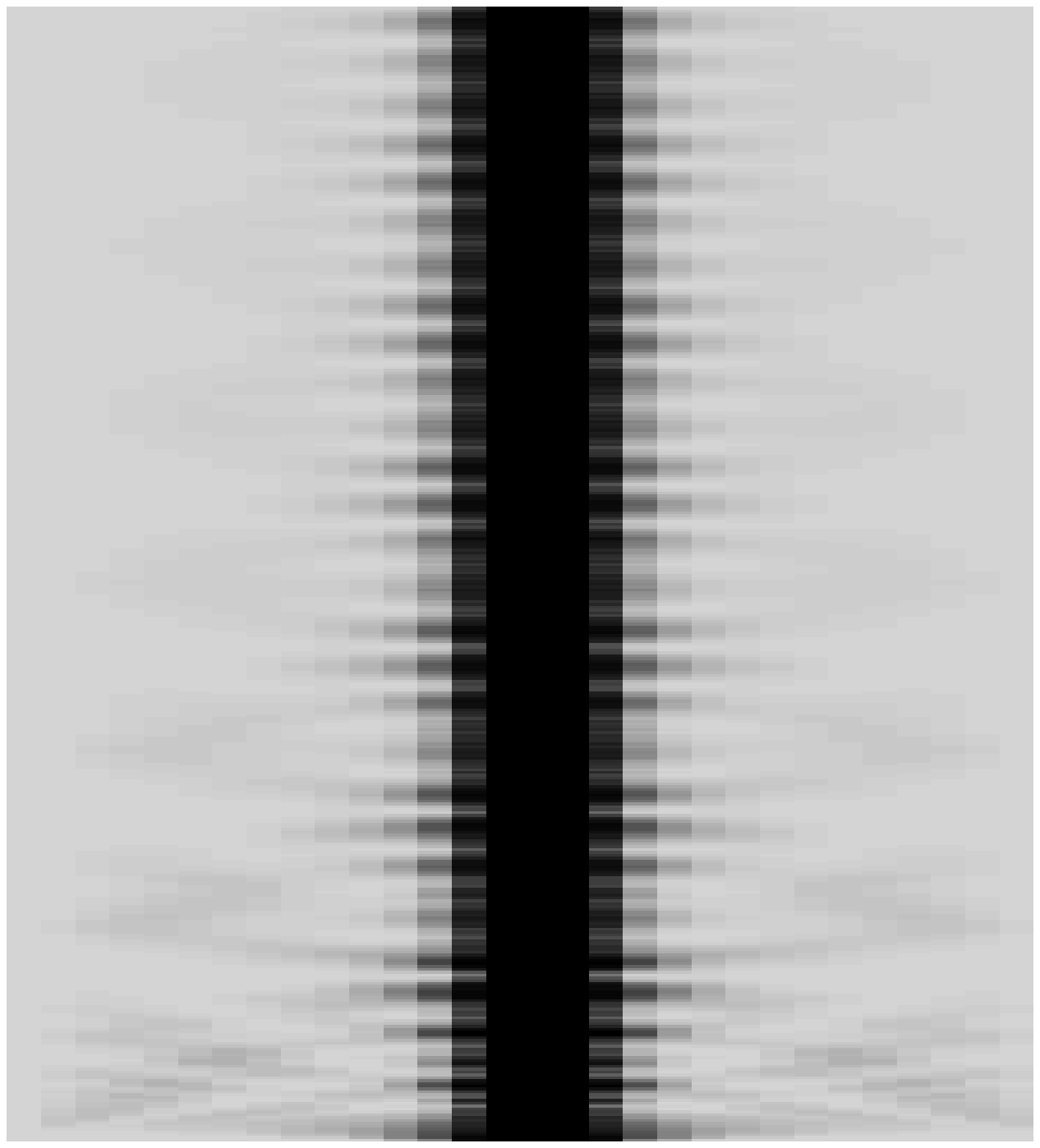}
\epsfxsize = 2.in
\epsffile{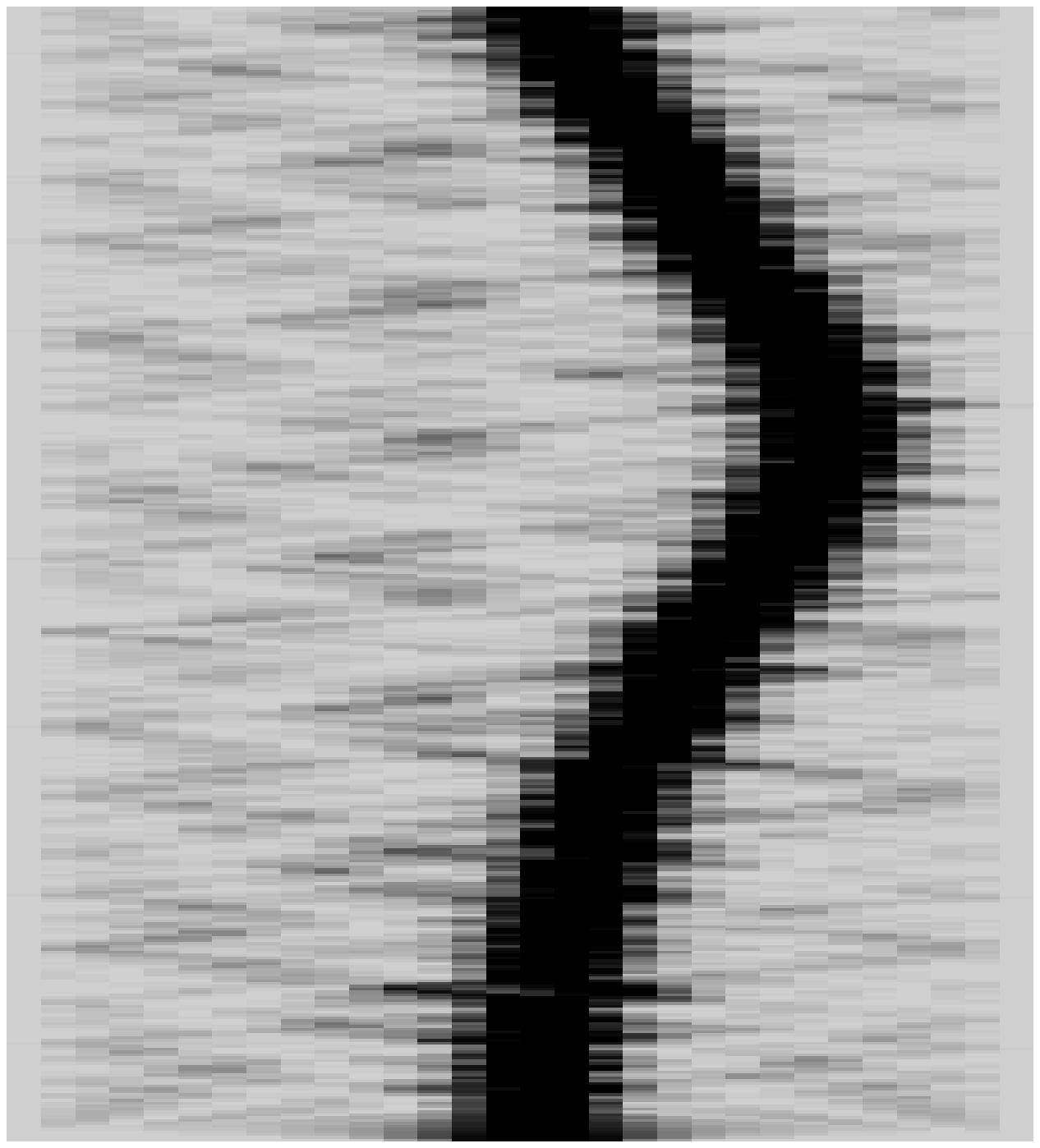}
\end{center}
\caption
{Energy landscapes of 31-site arrays. 
The injected odd-parity breather of amplitude
$A=0.5$ is initially centered
at site 15.  Time advances along the $y$-axis until $t=2000$.
A gray scale is used to represent the local energy,
with darker shading corresponding to more energetic regions. First
panel:
$\gamma=0.01$.  Second panel: $\gamma=1$.  Third panel: $\gamma=100$.}
\label{figure10}
\end{figure}

The evolution of the breather depends in an interesting way on
its initial location and on the damping parameter $\gamma$ when the latter
is either very small or very large.  Figure~\ref{figure10} shows the early
evolution (up to $t=2000$) of an initially  slightly off-center breather
(site 15 of a 31-site chain), for
three values of the damping parameter.  The middle panel is for $\gamma=1$,
the damping we have considered so far.  The behavior of the excitation
in this panel
starts out as we have described it, that is, it sheds some energy
(medium--gray scale) that dissipates quickly.  Although a small fraction of
the energy that has been shed returns toward the breather,
it is neither sufficient nor of the long wavelength variety
to set it in motion; most of the shed energy simply 
dissipates into the zero-temperature bath.
The evolution of the breather proceeds much like that of a
breather initially centered in the middle of the chain (site 16), with only
a small modification of its decay time. 
This behavior is fairly robust for
values of $\gamma$ within an order of magnitude on either side of
$\gamma=1$ and for breathers that are excited not too near the chain
ends. 

The situation is rather different if $\gamma$ is either very small
(first panel) or very large (third panel).  The qualitative
similarity between these two extreme cases is apparent, and consistent
with our discussion of high and low damping similarities in a purely
harmonic chain; the chain
ends no longer effectively dissipate the energy that has been
shed by the breather, and so it returns to
perturb the breather and set it in motion.  In turn, this causes
the breather to decay more rapidly into more rapidly dissipated
lower-energy excitations~\cite{weasymp}.  In the very low damping case,
energy that arrives at the chain ends can not go anywhere except
back, much like a whip.  In the very high $\gamma$ case the end sites
are so damped that they can absorb very little energy from the rest of the
chain, much like a wall.  We have followed these particular histories over
our usual time span of 3 million time units and find the
decay times $\tau= {\mathcal O}(10^9)$
for $\gamma=1$, ${\mathcal O}(10^6)$ for $\gamma=0.01$, and
${\mathcal O}(3\times 10^6)$ for $\gamma=100$.  
The specific values change depending on the initial location of the
breather and the values of the other parameters of the system, but the
trend is clear.

An odd parity breather initially centered {\em exactly}
in the middle of the chain constitutes a singular case
when damping is very low or very high,
with relative decay rates {\em opposite} to those
reported above.  While the $\gamma\sim 1$ results are not much affected
by the initial location of the excitation (provided it is
far from the chain
ends), in this peculiar case the extreme--$\gamma$ cases lead to
{\em slower} decay than for $\gamma\sim 1$.
In this uniquely symmetric case, the breather is
perturbed from both sides by {\em identical} energy pulses
that return from the ends of the chain.  In the absence of symmetry
breaking, the breather is therefore not set in motion,
and instead simply re-absorbs this energy (and re-emits and re-absorbs
energy in increasingly smaller amounts) .  Since the energy that returns
from the chain ends is greater in the extreme $\gamma$ cases than it is
for intermediate $\gamma$, the chain energy remains higher, and 
the decay is thus slower.  

Breather decay times are strongly dependent on chain length:
the decay times increase markedly, as does the total lifetime of the
breather, with increasing $N$ because finite size effects and disturbances
scattered back from chain boundaries are reduced.  This is already apparent
when one compares the $N=31$ and $N=21$
results in Fig.~\ref{figure9}.  Whereas a breather of initial amplitude
$A=0.5$ created at the center of a 31-site chain has barely decayed over 3
million time units, a breather of the same initial amplitude in a 21-site
chain has disintegrated completely well before that.  With $A=0.5$ and
$\gamma=1$ for the centered breather we find 
$\tau=2.8\times 10^9$ for $N=31$ (as reported above),
$\tau=3.2\times 10^{12}$ for $N=41$, and $\tau=3.6\times 10^{15}$ for
$N=51$.  

Exponential decay points to a single rate-limiting decay channel
for the energy.  This channel is the shedding of energy in the
form of phonons and/or
lower energy localized excitations by the
breather.  The degradation of lower-energy localized excitations,
and the dissipation of energy into the zero-temperature
bath, are much faster processes.  However, the relaxation rate associated
with the shedding process is strongly dependent on chain length, breather
location, and other system parameters.

To tie together all the scenarios that we have presented in support of our
picture of breather dynamics in mixed arrays, we add one more
``experiment": we follow the dynamics of a
breather injected into a relaxing chain {\em after} the long
wavelength phonons
have decayed, but before the thermal relaxation process is complete.  
If our picture is correct, the breather lifetime should be much longer
than that of one injected at time $t=0$ (albeit perhaps shorter than that
of the same breather injected in a zero temperature chain).  We do indeed
find that the breather stability improves dramatically.
For example, for a zero-temperature injected
breather of initial amplitude $A=0.6$ in a chain of 31 sites
we reported above that over 3 million time units the relaxation time of
the breather is $\tau=2.8\times10^{14}$. In a chain
initially thermalized at $T=10^{-5}$ and then allowed to relax, if we wait
until $t=15,000$ before injecting the same breather we find a somewhat
shortened but still very long decay time
of $\tau=5.9\times10^{13}$, in any case much longer than it would
be if injected at $t=0$. 

We end this section with a caveat: all the exponential and
quasi-exponential slow decays reported for the various scenarios are for
{\em single} realizations.  In thermalized scenarios where breathers
are created spontaneously (but not necessarily in every realization), an
ensemble average could lead to a time dependence of the array energy
that may be complicated
by the occurrence of a broad range of relaxation times.  In the other
scenarios, e.g. where breathers are injected ``manually," a range of
relaxation times might also occur in an ensemble if the location of the
breather varies from one realization to another.

\section{Conclusions}
\label{concl}

We have studied the dynamics and relaxation of breathers in
Fermi-Pasta-Ulam arrays whose boundaries are connected
through damping terms to a
zero temperature heat bath.  We find that 
breather dynamics and relaxation in these nonlinear arrays with quartic
inter-particle interactions proceed along energetic pathways 
that are highly sensitive to the 
presence or absence of quadratic contributions to the interactions.

To understand the role of quadratic interactions we have recalled that
phonons in these arrays relax independently of one another
(provided the damping at the
boundaries is not too strong), that the phonon relaxation time is
wavevector dependent, and that phonons therefore relax sequentially,
starting with the longest wavelengths for the free-end boundary conditions
used in our work.  We have also pointed out that breathers are
fragile against collisions with long wavelength phonons and also
with other localized
nonlinear excitations.  Breathers are therefore quite robust in the
absence of long wavelength phonons and of other nonlinear excitations, but
are rapidly degraded in the presence of either. To arrive at these
conclusions, and to investigate them more quantitatively (at least
numerically), we have performed a number of numerical experiments
involving the spontaneous and the manual creation of breathers in arrays
initially at finite temperatures and at zero temperature.  

Breather decay brought about by collisions with long wavelength phonons and
with other nonlinear excitations, and by the associated breather
motion, is rapid, even more rapid than typical relaxation times of high
frequency phonons.  The actual process of breather disintegration caused by
collisions and associated motion is one whereby the breather breaks up
rapidly into lower energy excitations.  These mechanisms of breather decay
cause their lifetimes to be short in systems that contain such
excitations.  Examples include  thermalized purely anharmonic arrays that
have no efficient way to eliminate
their thermal excitations.  Even at zero
temperature, a manually injected
localized mode in a purely anharmonic array will (perhaps after a prolonged
period of stability) eventually succumb rather suddenly and rapidly to
the very perturbations produced during the relaxation process as the
localized mode re-accommodates itself and/or the energy it sheds 
is reflected back by the system boundaries
(finite size effects). Breather decay is also rapid if a breather is
manually injected in a thermalized mixed array, mainly due to the effects
of long wavelength phonons. Since these phonons are
highly destructive
of breathers, breather degradation is observed even when
the temperatures involved
are extremely low.  

On the other hand, breathers that are isolated from the effects of long
wavelength phonons and of other nonlinear excitations persist for extremely
long times. Examples are spontaneous breathers that arise during thermal
relaxation of a mixed array.  Long wavelength phonons as well as
other nonlinear excitations that themselves decay into phonons
are the first to relax, and spontaneous
breathers make their appearance when the system has already been swept
clean of these particular excitations.
Short wavelength phonons do not destroy breathers; on the contrary, they
tend to be absorbed by them and thus to contribute to their stability.  
The crucial importance of phonons and their ability to relax into the cold
temperature heat bath (especially the long wavelength phonons) is thus
evident: spontaneously created breathers in mixed arrays can persist
because of the phonon dynamics, whereas the absence of phonons (sonic
vacuum) and the consequent difficulty of purely anharmonic arrays to 
eliminate ``offending" excitations leads to the inability of
spontaneously created breathers to persist.   For the same reasons,
manually injected breathers in mixed arrays can persist for a very long
time if inserted in a zero temperature array, or in an array in which long
wavelength phonons and other nonlinear excitations have already relaxed,
but breathers will not survive if injected in a thermalized mixed chain,
or even in a chain that is allowed to cool down to a very low but nonzero
temperature, no matter how low the temperature.

Having established conditions that favor breather longevity (mixed
anharmonic chains at zero temperature or partially relaxed to zero
temperature), we have 
noted that even these breathers must eventually relax through some
intrinsic energy shedding process since the chain must eventually
equilibrate to zero temperature.  This intrinsic process is very slow and
essentially exponential over very long periods of time, although some
deviation from strict exponentiality is eventually observed because the
relaxation time is amplitude dependent.  Thus, as the breather slowly loses
energy/amplitude, this ``single" relaxation time necessarily decreases.
This quasi-exponential decay continues until the breather frequency
(which decreases with decreasing amplitude) approaches the phonon band,
at which point the breather quickly breaks up into phonons that decay
rapidly.  The very slow quasi-exponential decay is indicative of
essentially a single leakage channel. We have observed that the slow
decay rates are dependent on system parameters, on breather
location, and on breather amplitude. Therefore, whereas single observations
of energy relaxation of systems supporting one long-lived breather will
lead to essentially exponential decay over many decades of time, ensemble
averages might show more complex behavior.  This is a specially strong
caveat in experiments involving spontaneous breather creation and the
associated possibility of realizations in which no breathers occur at all.

\section*{Acknowledgments}
This work was supported by the Engineering Research Program of
the Office of Basic Energy Sciences at the U. S. Department of Energy
under Grant No. DE-FG03-86ER13606. Support was also
provided by a grant from the University of California Institute for
M\'exico and the United States (UC MEXUS) and the Consejo Nacional de
Ciencia y Tecnolog\'{\i}a de M\'{e}xico (CoNaCyT).


\begin{thebibliography}{99}
\bibitem{thermalization} Our thermalization methods, which involve adding
Langevin forces to all the equations of motion, are detailed in our
earlier work,~\cite{we1d,we2d,weasymp}.
\end{thebibliography}
\end{document}